\documentclass{aa}

\usepackage{graphicx}
\usepackage{color}
\usepackage{txfonts}

\newcommand{\beq}{\begin{equation}}
\newcommand{\eeq}{\end{equation}}
\newcommand{\beqa}{\begin{eqnarray}}
\newcommand{\eeqa}{\end{eqnarray}}

\begin{document}

\title
{
Torsional oscillations of longitudinally inhomogeneous coronal loops
}

\subtitle{}

\author{
      T.V. Zaqarashvili        \inst{1}
      \&
      K. Murawski              \inst{2}
        }

\offprints{T. Zaqarashvili \email{temury@genao.org}}

\institute{
Georgian National Astrophysical Observatory (Abastumani Astrophysical Observatory),
Kazbegi Ave. 2a, Tbilisi 0160, Georgia
\and
Group of Astrophysics and Gravity Theory,
Institute of Physics, UMCS, ul. Radziszewskiego 10, 20-031 Lublin, Poland
  }

\date{received / accepted }

\abstract {}{We explore the effect of an inhomogeneous mass density
field on frequencies and wave profiles of torsional Alfv\'en
oscillations in solar coronal loops.}{Dispersion relations for
torsional oscillations are derived analytically in limits of weak
and strong inhomogeneities. These analytical results are verified by
numerical solutions, which are valid for a wide range of
inhomogeneity strength.}{It is shown that the inhomogeneous mass
density field leads to the reduction of a wave frequency of
torsional oscillations, in comparison to that of estimated from mass
density at the loop apex. This frequency reduction results from the
decrease of an average Alfv\'en speed as far as the inhomogeneous
loop is denser at its footpoints. The derived dispersion relations
and wave profiles are important for potential observations of
torsional oscillations which result in periodic variations of
spectral line widths.}{Torsional oscillations offer an additional
powerful tool for a development of coronal seismology.}

\titlerunning{Torsional oscillations of a coronal loop}
\authorrunning{T. Zaqarashvili \& K. Murawski}


\maketitle

\section{Introduction}
Recent space-based observations revealed a presence of various kinds
of magnetohydrodynamic (MHD) waves and oscillations in the solar
corona. These observations as well as modeling of MHD waves are
important as these waves contribute to the coronal heating problem
(Roberts \cite{roberts}) and they may consist unique tool of a
coronal seismology (Edwin \& Roberts \cite{edwin83}, Nakariakov \&
Ofman \cite{nakariakov2}). Fast kink (Aschwanden et al.
\cite{aschwanden}, Nakariakov et al. \cite{nakariakov1}, Wang \&
Solanki \cite{wang2}) and sausage (Nakariakov 2003, Pascoe et al.
\cite{pascoe})
as well as slow (de Moortel et al. 2002, Wang et al. 2003)
magnetosonic oscillations were observed to be associated
either with or without a solar flare. Analytical studies of these
oscillations in coronal loops were carried on over the last few
decades, amongst others, by Edwin \& Roberts (1982, 1983), Poedts \&
Boynton (1996), Nakariakov (2003), Van Doorsselaere et al.
(2004a,b), Ofman (2005), Verwichte et al. (2006) and Di\'az et al.
(2006).

Coronal loops act as natural wave guides for magnetosonic and
torsional Alfv\'en waves. The later are purely azimuthal
oscillations in cylindrical geometry. In the linear regime, Alfv\'en
oscillations do not lead to mass density perturbations. As a result,
contrary to magnetosonic waves, torsional Alfv\'en waves can be
observed only spectroscopically.
While propagating from the base of the solar corona along open
magnetic field lines, these waves may lead to an increase of a
spectral line width with height (Hassler et al. \cite{has}, Banerjee
et al. \cite{banerjee}, Doyle et al. \cite{doyle}). In closed
magnetic field structures, such as coronal loops, these waves can be
observed indirectly as periodic variations of non-thermal broadening
of spectral lines (Zaqarashvili \cite{zaqarashvili}).

Alongside magnetosonic waves, torsional oscillations can be used to
infer, in the framework of coronal seismology, plasma properties
inside oscillating loops. These oscillations are an ideal tool of
coronal seismology as their phase speed depends alone on plasma
quantities within the loop, while wave speeds of magnetosonic
oscillations are influenced by plasma conditions in the ambient
medium. Having known mass density within a loop, coronal seismology,
that is based on torsional oscillations, enables to estimate a
magnetic field strength. Torsional oscillations are potentially
important in the context of rapid attenuation of coronal loop kink
oscillations (Aschwanden et al. \cite{aschwanden}, Nakariakov et al.
\cite{nakariakov1}). One of a few suggested mechanisms of the
attenuation is a resonant absorption of fast magnetosonic kink waves
by azimuthal Alfv\'en waves (Ruderman \& Roberts 2002). This process
may lead to a formation of torsional oscillations in the outer part
of a loop. As a result, spotting torsional oscillations after the
kink mode was attenuated would serve as an evidence of this
attenuation mechanism.

A theoretical study of Alfv\'en oscillations in a coronal loop was
carried on recently by Gruszecki et al. (2007) who considered
impulsively generated oscillations in two-dimensional straight and
curved magnetic field topologies. They found that lateral leakage of
Alfv\'en waves into the ambient corona is negligibly small. However,
mass density profiles were adopted homogeneous within the loop,
while the real conditions there are much more complex.

Despite of significant achievements in a development of realistic
models there is still much more effort required to develop our
knowledge of wave phenomena in coronal loops. A goal of this paper
is to study the influence of inhomogeneous mass density fields on
spectrum of torsional oscillations. The paper is organized as
follows. Analytical solutions for torsional oscillations in a
longitudinally inhomogeneous coronal loop are presented in Sect.~2.
The numerical results are showed in Sect.~\ref{sec:num_res}.
Guidelines for potential observations of these oscillations are
presented in Sect.~4. This paper is concluded by a discussion and a
short summary of the main results in Sect.~\ref{sec:sum}.
\section{Analytical model of torsional oscillations}
We consider a coronal loop of its inhomogeneous mass density
${\varrho}_{\rm 0}(z)$ and length $2L$, that is embedded in a
uniform magnetic field ${\bf B}=B_{\rm 0}{\bf\hat z}$. Small
amplitude torsional Alfv{\'e}n waves in a cylindrical coordinate
system ($r$, $\phi$, $z$), in which plasma profiles depend on a
longitudinal coordinate $z$ only, can be described by the following
linear equations:
\beqa
\label{eq:2eqs1}
{{\partial u_{\phi}}\over {\partial t}}&=&{{B_{\rm 0}}\over
{4\pi{\varrho}_{\rm 0}(z)}}{{\partial b_{\phi}}\over {\partial z}}\, , \\
{{\partial b_{\phi}}\over {\partial t}}&=&B_{\rm 0}{{\partial
u_{\phi}}\over {\partial z}}\, ,\label{eq:2eqs2} \eeqa
where $u_{\phi}$ and $b_{\phi}$ are the velocity and magnetic field
components of Alfv{\'e}n waves.

\begin{figure}[!t]
\begin{center}
\includegraphics[scale=0.5]{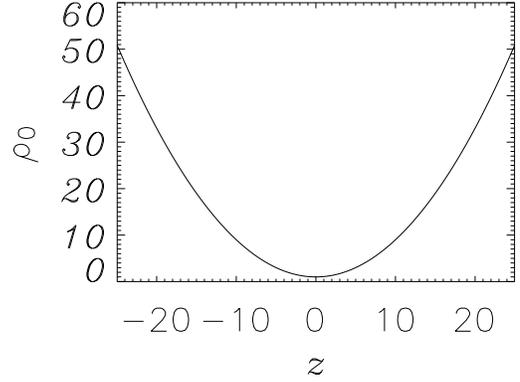}
\caption{\small Spatial profile of the background mass density,
${\varrho}_{\rm 0}(z)$, given by Eq.~(\ref{eq:rho})
with $\alpha^2=50$. The mass density and length are expressed in
units of $10^{-12}$ kg m$^{\rm -3}$ and 1 Mm, respectively. }
\label{fig:bgd}
\end{center}
\end{figure}

These equations can be easily cast into a single wave equation
\beq {{\partial^2 u_{\phi}}\over {\partial z^2}} - {1 \over V^2_{\rm
A}(z)}{{\partial^2 u_{\phi}}\over {\partial t^2}}=0\,, \eeq
where $V_A(z)=B_{\rm 0}/\sqrt{4\pi{\varrho}_{\rm 0}(z)}$ is the
Alfv{\'e}n speed.
Assuming that $u_{\phi} \sim \exp(i\omega t)$, where $\omega$ is a
wave frequency, we get the equation
\beq
{{\partial^2 u_{\phi}}\over {\partial z^2}} + {\omega^2 \over V^2_A}u_{\phi}=0\, .
\label{eq:u_phi}
\eeq
For a trapped solution $u_{\phi}$ must satisfy line-tying boundary
conditions which are implemented by setting
\beq u_{\phi}(z=\pm L)=0\, . \label{eq:bc} \eeq
Equation~(\ref{eq:u_phi}) with condition (\ref{eq:bc}) consists the
well-known Sturm-Liuville problem which solution depends on the
profile of $V_{\rm A}(z)$. We model the coronal loop by a rarefied
plasma at the loop apex (at $z=0$) and by a compressed plasma at the
loop footpoints ($z=\pm L$). Specifically, we adopt
%
\beq
{\varrho}_{\rm 0}(z)={\varrho}_{\rm 00}\left (1+{\alpha^2}{z^2\over L^2}\right )\, ,
\label{eq:rho}
\eeq
where ${\varrho}_{00}$ is the mass density at the loop apex and
$\alpha^2$ is a parameter which defines a strength of the
inhomogeneity. For $\alpha^2=0$ the above mass density profile
corresponds to a homogeneous loop, while for a larger value of
$\alpha^2$ the medium is more inhomogeneous. Figure~\ref{fig:bgd}
illustrates
${\varrho}_{\rm 0}(z)$
for $\alpha^2=50$. The mass density is described by
Eq.~(\ref{eq:rho}) with ${\varrho}_{\rm 00}=10^{-12}$ kg m$^{\rm
-3}$ and $L=25$ Mm. Note that plasma is compressed at $z=\pm L$.
%
Substituting Eq.~(\ref{eq:rho}) into Eq.~(\ref{eq:u_phi}), we obtain
\beq
{{\partial^2 u_{\phi}}\over {\partial z^2}} + {\omega^2 \over
V^2_{A0}}\left (1+{\alpha^2}{z^2\over L^2}\right )u_{\phi}=0\,,
\label{eq:u_prof}
\eeq
where $V_{\rm A0}=B_{\rm 0}/\sqrt{4\pi{\varrho}_{\rm 00}}$.
With a use of the notation
\beq
y \equiv u_{\phi}, \hspace{3mm}
x \equiv \sqrt{{2\alpha\omega}\over {V_{\rm A0}L}}z, \hspace{3mm}
a \equiv -{\omega \over V_{\rm A0}}{L\over {2\alpha}}
\eeq
%
Eq.~(\ref{eq:u_prof}) can be rewritten in the form of Weber
(parabolic cylinder) equation  (Abramowitz \& Stegun
\cite{abramowitz})
\beq
{{\partial^2 y}\over {\partial x^2}} + \left({x^2\over 4}-a \right)y = 0\, .
\label{eq:weber}
\eeq
Standard solutions to this equation are called Weber (parabolic
cylinder) functions (Abramowitz \& Stegun \cite{abramowitz})
\beq W(a,\pm x)={({\cosh {\pi a})^{1/4}}\over
{2\sqrt{\pi}}}\left(G_{\rm 1} y_{\rm 1}(x) \mp {\sqrt {2}G_{\rm
3}y_{\rm 2}(x)}\right)\,, \eeq
where
\beq
G_{\rm 1} = \left\vert \Gamma\left({1\over 4}+{{ia}\over 2}\right)\right\vert,\hspace{3mm}
G_{\rm 3} = \left\vert\Gamma\left ({3\over 4}+{{ia}\over 2}\right )\right\vert
\eeq
and $y_{\rm 1}(x),\,\,y_{\rm 2}(x)$ are respectively even and odd
solutions to Eq.~(\ref{eq:weber})
$$ y_{\rm 1}(x)=1+a{{x^2}\over {2!}}+\left (a^2-{1\over 2}\right ){{x^4}\over
{4!}}+\cdots,
$$
$$
\, y_{\rm 2}(x)=x+a{{x^3}\over {3!}}+\left (a^2-{3\over 2}\right
){{x^5}\over {5!}}+\cdots.
$$

\subsection{Two limiting solutions}
Periodic solutions to Eq.~(\ref{eq:weber}) can be written
analytically in the limiting cases: (a) for a large value of $a$ but
a moderate value of $x$; (b) for a large $x$ but a moderate $a$. The
first (second) case corresponds to $\alpha^2 \ll 1$ ($\alpha^2 \gg
1$).
\subsubsection{Weakly inhomogeneous plasma}
We consider first the case of a weakly inhomogeneous mass density
field, i.e. $\alpha^2 \ll 1$. In this case we have
\beq
a < 0,\,\,\,-a\gg x^2,\,\,\,p\equiv\sqrt{-a}\, .
\eeq
%
We adopt the following expansion (Abramowitz \& Stegun
\cite{abramowitz}):
\beqa
\nonumber
W(a,x)+iW(a,-x)=\\
\sqrt{2}W(a,0)\exp{[v_r + i(p x + {\pi/4} +v_i)]}\,,
\label{eq:expand}
\eeqa
where
\beqa
W(a,0)    = {1\over 2^{3/4}}\sqrt{{G_{\rm 1}}\over {G_{\rm 3}}}\,,\\
v_{\rm r} = -{{(x/2)^2}\over {(2p)^2}}+{{2(x/2)^4}\over {(2p)^4}}+\cdots \,,\hspace{3mm}
v_{\rm i} = {{2/3(x/2)^3}\over {2p}} + \cdots \,.
\eeqa
As a result of relation $-a\gg x^2$ we have from Eq.~(\ref{eq:expand})
\beqa
W(a,x)=\sqrt{2}W(a,0)\exp{\left (-{{x^2}\over {16p^2}}\right
)}\cos{\zeta}\, , \\
W(a,-x)=\sqrt{2}W(a,0)\exp{\left (-{{x^2}\over {16p^2}}\right
)}\sin{\zeta}\, ,\\
\zeta\equiv p x + {\pi/4} +{{x^3}\over {24p}}\, . \eeqa
The general solution to Eq.~(\ref{eq:weber}) is
\beq
u_{\phi}=c_{\rm 1}W(a,x)+c_{\rm 2}W(a,-x)\,,
\label{eq:gen}
\eeq
where $c_{\rm 1}$ and $c_{\rm 2}$ are constants.

For a homogeneous loop, i.e. ${\alpha}^2 = 0$, we recognize the well
known solution
\beq u_{\phi}\sim c_{\rm 1}\cos{\left (k z + {\pi/4}\right )} +
c_2\sin{\left (k z + {\pi/4}\right )}. \label{eq:u_phi_p} \eeq
Here wave number $k$ satisfies the following homogeneous dispersion
relation:
\beq
k={\omega \over V_{A0}}\,.
\eeq
Line-tying boundary conditions of Eq.~(\ref{eq:bc}) lead then
%
%
to discrete values of the wave frequency, viz.
\beq {\omega_n}={{n\pi}\over 2L}{{V_{A0}}\over
{1+{{\alpha^2}/6}}},\hspace{3mm} n=1,2,3,\ldots\, .
\label{eq:omega_small} \eeq
From this dispersion relation we infer that in a comparison to the
loop with a homogeneous mass density distribution, ${\varrho}_{\rm
00}$, the weakly inhomogeneous mass density field results in a
decrease of a wave frequency. This reduction is a consequence of the
fact that the inhomogeneous loop is denser at its footpoints, so the
average Alfv\'en speed is decreased. To show this, we compare the
results for the inhomogeneous loop with the homogeneous loop with
the same average density, so that both loops contain exactly the
same mass (Andries et al. \cite{andries}). We introduce a frequency
difference
\beq\label{eq:delta_omega} \Delta{\omega}_{\rm n} = {\omega}_{\rm n}
- \bar{\omega}_{n}\,, \eeq
where
\beq\label{eq:bar_omega} \bar{\omega}_{\rm n} = \frac{n\pi}{2L}\bar
V_{\rm A0}= \frac{n\pi}{2L} \frac{B_{\rm 0}}{\sqrt{4\pi
\bar\varrho_0}} \eeq
corresponds to the average mass density
\beq\label{eq:bar_rho} \bar\varrho_0 = \frac{1}{2L} \int_{-L}^{L}
{\varrho}_{\rm 0}(z) dz = {\varrho}_{\rm 00} \left
(1+\frac{\alpha^2}{3}\right )\,. \eeq
Substituting Eq.~(\ref{eq:bar_rho}) into Eq.~(\ref{eq:bar_omega}),
we obtain
\beq \bar{\omega}_{\rm n} = \frac{n\pi}{2L} \frac{V_{\rm
A0}}{\sqrt{1+{\alpha^2}/{3}}}\,. \eeq
From Eqs.~(\ref{eq:delta_omega}) and (26) we find that
$\Delta{\omega}_{\rm n}\leq 0$. Here we infer that in comparison to
the average mass density case the wave frequency is reduced, but as
a result of $\alpha^2 \ll 1$ the frequency reduction is small. This
is in a disagreement with Fermat's law and with the results of
Murawski et al. (\cite{murawski}) who showed that sound waves
experience frequency increase in a case of a space-dependent random
mass density field.

\subsubsection{Strongly inhomogeneous plasma}
We discuss now a strongly inhomogeneous mass density case, i.e.
$\alpha^2 \gg 1$. This case corresponds to $x\gg \vert a\vert$. In
this limit we get (Abramowitz \& Stegun \cite{abramowitz})
\beqa
W(a,x)  &=& \sqrt{2k/x}(s_{\rm 1}(a,x)\cos(\xi)-s_{\rm 2}(a,x)\sin(\xi))\, ,\\
W(a,-x) &=& \sqrt{2/kx}(s_{\rm 1}(a,x)\sin(\xi)-s_{\rm
2}(a,x)\cos(\xi))\, , \eeqa
where
\beqa \xi \equiv \frac{x^2}{4} - a \ln{x} + \frac{\pi}{4} +
\frac{\arg \Gamma(1/2 +ia)}{2}\,,\\
k = \sqrt{1+e^{2\pi a}}-e^{\pi a},\, \\
s_{\rm 1}(a,x) \sim 1 + {v_{\rm 2}\over {1!2x^2}}-{u_4\over {2!2^2x^4}}- \cdots \,, \\
s_{\rm 2}(a,x) \sim -{u_{\rm 2}\over {1!2x^2}}-{v_4\over
{2!2^2x^4}}+\cdots \eeqa
with
\beq u_r+iv_r=\Gamma(r+1/2+ia)/\Gamma(1/2+ia),\,\,\, r=2,4, \ldots\,
.\eeq
%

%
%

The boundary conditions of Eq.~(\ref{eq:bc}) lead to the discrete
frequency spectrum
\beq
\omega_n={{n\pi}\over {\alpha}}{V_{A0}\over L}\, .
\label{eq:omega_big}
\eeq

Here we infer that the strongly inhomogeneous mass density field
results in a significant decrease of a wave frequency in comparison
to the case of the loop with the constant density, ${\varrho}_{\rm
00}$. This wave frequency decrease is a consequence of the fact that
the inhomogeneous loop is denser at its footpoints. Substituting
Eq.~(34) into Eq.~(\ref{eq:delta_omega}) we find that
$\Delta{\omega}_{\rm n}> 0$. This wave frequency decrease, in a
comparison to the case of an average mass density is now in an
agreement with Fermat's law and with the results of Murawski et al.
(2004).
\section{Numerical results}\label{sec:num_res}
Numerical simulations are performed for Eqs.~(\ref{eq:2eqs1}),
(\ref{eq:2eqs2}) with an adaptation of CLAWPACK which is a software
package designed to compute numerical solutions to hyperbolic
partial differential equations using a wave propagation approach
(LeVeque 2002). The simulation region $(-L,L)$ is covered by an
uniform grid of $600$ numerical cells. We verified by convergence
studies that this grid does not introduce much numerical diffusion
and as a result it represents well the simulation region. We set
reflecting boundary conditions at the left and right boundaries of
the simulation region.

%
\begin{figure}[!h]
\begin{center}
\includegraphics[scale=0.45]{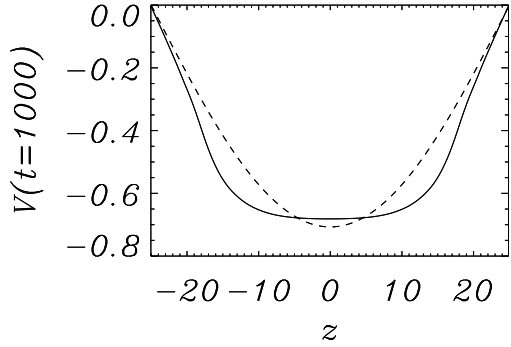}
\caption{\small Numerically evaluated velocity profile $u_\phi$ at
$t=1000$ s for $\alpha^2=50$ (solid line). This profile corresponds
to the mode number $n=1$. Note that as a result of strong
inhomogeneity, $u_\phi$ departs from the sine-wave which corresponds
to $\alpha^2=0$. The dashed line corresponds to Eq. (20) with
$c_1=c_2=0.5$.} \label{fig:V_z}
\end{center}
\end{figure}
Figure~\ref{fig:V_z} shows a spatial profile of velocity
$u_{\phi}(z)$ for $\alpha^2=50$, drawn at $t=1000$ s (solid line).
This spatial profile results from the initial Gaussian pulse that
was launched at $t=0$ in the center of the simulation region, at
$z=0$. It is noteworthy that the sine-wave profile of
Eq.~(\ref{eq:u_phi_p}), which is valid for $\alpha^2=0$ (dashed
line), is distorted by the strong inhomogeneity which takes place
for the case of $\alpha^2=50$.
\begin{figure}[!h]
\begin{center}
\includegraphics[scale=0.45]{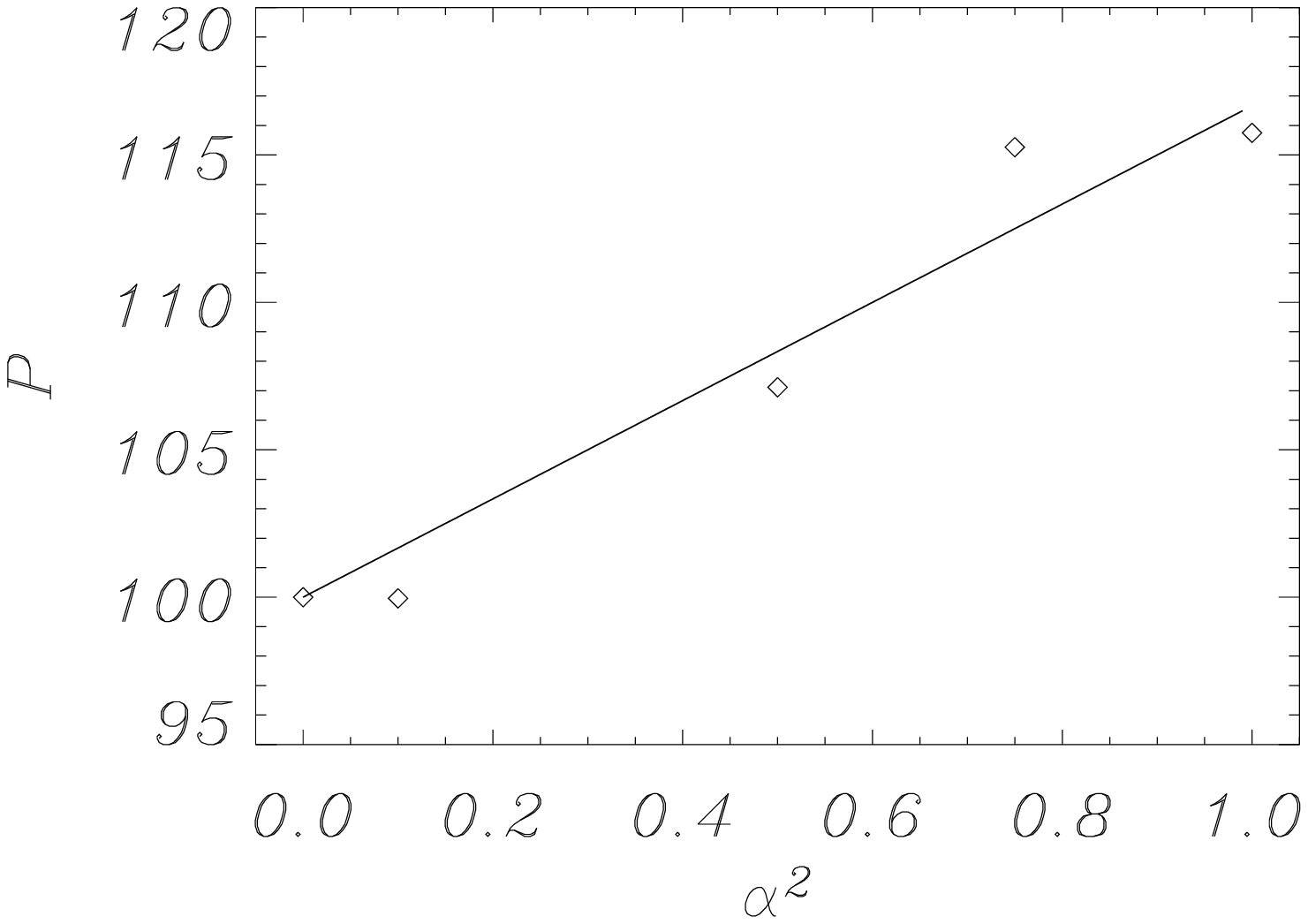}
\includegraphics[scale=0.45]{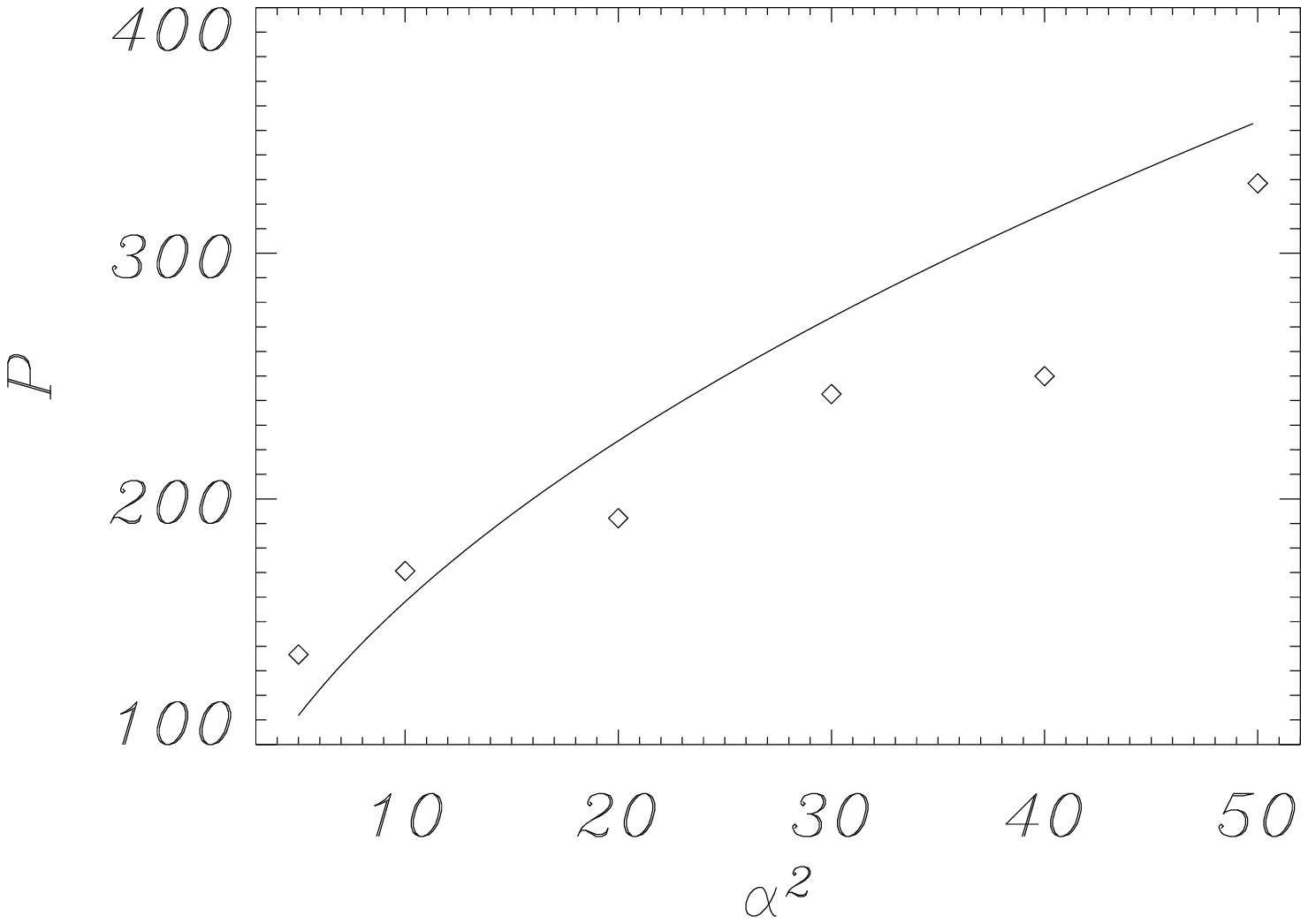}
\caption{\small
Wave period $P=\omega/2\pi$ vs. ${\alpha}^2$ for the mode number $n=1$. Diamonds correspond to
the numerical solutions to Eqs.~(\ref{eq:2eqs1}), (\ref{eq:2eqs2}). Solid lines are drawn with the use of
the analytical solution to Eqs.~(\ref{eq:omega_small}) and ~(\ref{eq:omega_big}). The wave period is expressed in seconds.
}
\label{fig:P_alpha2}
\end{center}
\end{figure}

As a consequence of the inhomogeneity wave period is altered.
Figure~\ref{fig:P_alpha2} displays wave period $P$ vs. inhomogeneity
parameter $\alpha^2$. Diamonds represent the numerical solutions
while the solid lines correspond to the analytical solution to
Eqs.~(\ref{eq:omega_small}) (top panel) and ~(\ref{eq:omega_big})
(bottom panel). Wave periods were obtained by Fourier analysis of
the wave signals that were collected in time at the fixed spatial
location, $z=0$. It is discernible that
the numerical data fits quite well to the analytical curves.
A growth of wave period $P$ with $\alpha^2$ results from wave scattering on centers of the inhomogeneity and
it can be explained on simple physical grounds. In an inhomogeneous field wave frequency ${\omega}_{\rm n}$
of the torsional oscillations can be estimated from the following formula:
\beq {\omega}_{\rm n} = \frac{n\pi}{2L} {\bar V_{\rm A0}}\, ,
\label{eq:omgn} \eeq
where ${\bar V_{\rm A0}}$ is the averaged Alfv\'en speed that is
expressed by Eq. (24). Using $P=2\pi/{\omega}_{\rm n}$ we obtain
\beq P=\frac{4L\sqrt{4\pi  {\bar \varrho}_{\rm 0}}} {nB_{\rm 0}}\, .
\eeq
As ${\bar \varrho}_{\rm 0}$ grows with $\alpha$, the growth of $P$
with $\alpha$ results in.
\section{Potential observations of torsional oscillations}
Torsional oscillations of a coronal loop may result in periodic
variations of spectral line non-thermal broadening (expressed by a
half line width, $\Delta\lambda_B$, hereafter HW) (Zaqarashvili
\cite{zaqarashvili}). For a homogeneous loop, HW can be expressed as
\begin{equation}
\Delta{\lambda}_B= {{uV_{\rm A0}{\lambda}}\over {c}}
{\left\vert{\sin}({\omega}_{\rm n} t){\sin}(k_{\rm n} z)\right\vert}\, ,
\end{equation}
where $u$ is an amplitude of oscillations, $\lambda$ is a wave
length of the spectral line and $c$ is the light speed. Periodic
variations of spectral line width depend on a height above the solar
surface: a strongest variation corresponds to the wave antinode and
the place of a lack of line width variation corresponds to the nodes
(loop footpoints). Therefore, time series of spectroscopic
observations may allow to determine a wave period. Knowing a length
of the loop, we may estimate the Alfv{\'e}n speed, which in turn
gives a possibility to infer the magnetic field strength in the
corona. We estimate the expected value of line width variations
which result from torsional oscillations. For a typical coronal
Alfv{\'e}n speed of $\sim$ 800 km/s, an amplitude of linear
torsional oscillation can be $\sim$ 40 km/s, which consists 5$\%$ of
the Alfv{\'e}n speed. For the "green" coronal line Fe XIV (5303
$\AA$) from Eq. (37) we obtain
\begin{equation}
\Delta{\lambda}_B \approx 0.7 \, {\AA}.
\end{equation}
This value is about twice larger than the original thermal
broadening of Fe XIV line. As a consequence, torsional oscillations
can be detected in time series of the green coronal line spectra.

For a weakly inhomogeneous distribution of mass density along a
loop, Eq.~(\ref{eq:omega_small}) enables to estimate the Alfv{\'e}n
speed at the loop apex with the help of the observed period of HW
variation and a loop length. For a strongly inhomogeneous density
profile along a loop, Eq. (\ref{eq:omega_big}) shows that a wave
period of torsional oscillations is not just the ratio of the loop
length to the Alfv{\'e}n speed, but it strongly depends on the rate
of inhomogeneity, $\alpha^2$. Therefore, an additional effort is
required in order to apply the method of coronal seismology for
torsional oscillations. A spatial variation of mass density along
the loop can be estimated by a direct measurement of spectral line
intensity variation along the loop. Then, the estimated variation
can be fitted to Eq.~(\ref{eq:rho}), and hence a value of $\alpha^2$
can be inferred. Eq. (\ref{eq:omega_big}) provides a value of
$V_{\rm A0}$ at the loop summit. Another possibility is to collect
time series of spectroscopic observations at different positions of
the loop. A spatial variation of line width along the loop may be
compared to the theoretical plot of $u_{\phi}$ (Fig. 2), which
enables to estimate $\alpha^2$ and consequently Alfv{\'e}n speed at
the loop apex (with a use of Eqs.~(\ref{eq:omega_small}) or
(\ref{eq:omega_big})).
\section{Discussion and summary}\label{sec:sum}
It is commonly believed that Alfv\'en waves are
generated in the solar interior either by convection (granulation,
supergranulation) or by any other kinds of plasma flow (differential
rotation, solar global oscillations). Due to their incompressible
nature, these waves may carry energy from the solar surface to the
solar corona and therefore they may significantly contribute to
coronal heating and solar wind acceleration. In closed magnetic
loops the Alfv\'en waves may set up the standing torsional
oscillations, while in opened magnetic structures these waves may
propagate up to the solar wind. As a result, observations of
Alfv\'en waves can be of vital importance to the problems of plasma
heating and particle acceleration.

The Alfv\'en waves that propagate along open magnetic field lines
may lead to a growth
of a spectral line width with height (Hassler et al. \cite{has},
Banerjee et al. \cite{banerjee}; Doyle et al. \cite{doyle}).
However, at some altitudes the spectral line width reveals a sudden
fall off
(Harrison et al. \cite{har}; O'Shea et al. \cite{osh1,osh2}). This
phenomenon was recently explained by resonant energy transfer into
acoustic waves (Zaqarashvili et al. \cite{zaqarashvili1}).

On the other hand, the photospheric motions may set up torsional
oscillations in closed magnetic loop systems, which can be observed
spectroscopically as periodic variations of spectral line width
(Zaqarashvili \cite{zaqarashvili}). As a result, the observation of
Alfv\'en waves can be used as an additional powerful tool of coronal
seismology; the observed period and loop mean length enables to
estimate the Alfv\'en speed within a loop, which in turn makes it
possible to infer a mean magnetic field strength.

Besides their photospheric origin, torsional Alfv\'en waves can be
generated in the solar corona in a process of
resonant absorption of the global oscillations (Ruderman \& Roberts
\cite{ruderman}, Goossens et al. \cite{goossens}, Andries et al.
\cite{andries}, Terradas et al. \cite{terradas}). These oscillations
may excite Alfv\'en waves in the outer inhomogeneous part of a loop,
leading to attenuation of global oscillations and amplification of
torsional oscillations. These Alfv\'en oscillations can be detected
as periodic variations of spectral line width. As a consequence,
observations of Alfv\'en waves can be a key for a determination of a
damping mechanism of the loop global oscillations.

Dynamics of torsional Alfv\'en waves in a homogeneous loop can be
easily solved. However, real coronal loops are longitudinally
inhomogeneous, which leads to alteration of wave dynamics (Arregui
et al. \cite{arregui,arregui1}, Van Doorsselaere et al. 2004a,b,
Donnelly et al. \cite{donnelly}, Dymova \& Ruderman \cite{dymova},
McEwan et al. \cite{mc}). Therefore, the dynamics of Alfv\'en waves
in longitudinally inhomogeneous coronal loops
must be understood in order to provide analytical basis for
potential observations of torsional oscillations.

In this paper we discussed by analytical and numerical means
evolution of torsional Alfv\'en waves in an inhomogeneous mass
density field. The analytical efforts resulted in dispersion
relations which were obtained for a specific choice of an
equilibrium mass density profile. These dispersion relations were
written explicitly for two limiting cases: (a) weekly inhomogeneous
and (b) strongly inhomogeneous mass density fields. From these
dispersion relations we inferred that the inhomogeneity results in a
wave frequency reduction in comparison to that of estimated at the
loop summit.
This analytical finding is supported by the numerical data which
reveals that frequency reduction takes place outside the region of
validity of the analytical approach. As a result of that we claim
that a reduction of wave frequency is ubiquitous for the
inhomogeneous mass density field we considered. This reduction is a
consequence of wave scattering on inhomogeneity centers and it
results from reduction of the average Alfv\'en speed within a
coronal loop.
%
This frequency reduction has important implications
as far as wave observations are concerned. The analytical formulae
can be used for estimation of coronal plasma parameters and
therefore torsional Alfv\'en waves consist an additional powerful
tool of coronal seismology.
%
%

{\it Acknowledgments:} The authors express their thanks to the
referee, Prof. S. Poedts, for his stimulating comments. The work of
T.Z. is supported by the grant of Georgian National Science
Foundation GNSF/ST06/4-098. A part of this paper is supported by the
ISSI International Programme "Waves in the Solar Corona".

\end{document}